\documentclass{optica-article}

\journal{opticajournal} % for journals or Optica Open

\articletype{Research Article}

\usepackage{lineno}
\linenumbers % Turn off line numbering for Optica Open preprint submissions.

\newcommand{\df}[1]{\frac{d#1}{dt}}

\usepackage{physics}

\begin{document}
\nolinenumbers

\title{Optimizing state transfer in a three-qubit array via quantum brachistochrone method}

\author{Kseniia S. Chernova,\authormark{1} Andrei A. Stepanenko,\authormark{2,1} and Maxim~A.~Gorlach\authormark{1,*}}

\address{\authormark{1}School of Physics and Engineering, ITMO University, Saint Petersburg 197101, Russia\\
\authormark{2}London Institute for Mathematical Sciences, Royal Institution, London, UK}

\email{\authormark{*}m.gorlach@metalab.ifmo.ru} %% email address is required; see note below about the corresponding author designation

% use {asbstract*} to suppress the copyright line. Copyright information will be added in production

\begin{abstract*} 
Quantum brachistochrone method has recently emerged as a technique allowing one to implement the desired unitary evolution operator in a physical system within the minimal time. Here, we apply this approach to the problem of time-optimal quantum state transfer in the array of three qubits with time-varying nearest-neighbor couplings and analytically derive the fastest protocol.
\end{abstract*}

%%%%%%%%%%%%%%%%%%%%%%%%%%  body  %%%%%%%%%%%%%%%%%%%%%%%%%%
\section{Introduction}

% \blue{The story about quantum brachistochrone method, brief review of the previous works and available analytical solutions, citations.}

Advancement of quantum technologies requires to prepare, process and store quantum states within the minimal possible time while keeping reasonably high fidelity. This motivates the interest to {\it quantum optimal control theory}~\cite{qocPontryagin} which aims to tailor time-varying Hamiltonian of the system to achieve the desired quantum state with the predefined constraints on the Hamiltonian.

The most straightforward approach to this problem is the adiabatic evolution of the Hamiltonian which  
transforms the initial set of the eigenstates into the desired one. In particular, adiabatic evolution allows to transfer the particle in the array via the so-called Thouless pump~\cite{thouless}. However, this requires extremely slow variation of the Hamiltonian, which is impractical.

%Quantum computing requiring controllable high-fidelity state transition in the quantum systems with noise increased interest in finding this optimal control, especially in the systems with noise. 
%Although the adiabatic transition is a common approach, it requires relatively slow evolution~\cite{thouless}, which significantly reduces fidelity due to noise.

%At the same time, finding control that provides both high fidelity and minimum transition time is extremely challenging.

This limitation can be overcome using advanced methods such as conter-adiabatic driving known also as shortcuts to adiabaticity~\cite{sta, stacounteradiabatic} and Pontryagin maximum principle~\cite{qocPontryagin}. These techniques allow  to achieve the fastest quantum evolution and approach quantum speed limit by numerical optimization. Being powerful numerical tools, these techniques are restricted by the chosen form of the control and do not provide straightforward access to analytical solutions.

%To overcome this limitation and to prepare the desired quantum state within the minimal time, several methods have been proposed. For instance, the counteradiabatic driving technique like shortcuts to adiabaticity~\cite{sta, stacounteradiabatic} was suggested to obtain high fidelity of quantum operation in finite time. At the same time, quantum optimal control theory~\cite{qocPontryagin} allows one to achieve the fastest quantum evolution and quantum speed limit by numerical optimization. 

%Note, that the problem of finding minimum time and maximum fidelity of quantum operations can be solved by the variational method and reduces to the search the form of the time-dependent Hamiltonian.

A recently suggested alternative is quantum brachistochrone method~\cite{Carlini2006} which recasts a bi-parametric search of the minimum evolution time along with the maximum fidelity as a variational problem. Originally formulated by Carlini {\it et al}~\cite{Carlini2006} variational problem to find time-optional evolution of quantum states and Hamiltonian for given initial and final conditions has further been generalized to the operator form to find the time-optimal realization of a target unitary operation~\cite{Carlini2007}. 
Using this technique, one can derive the control protocol for systems with Hermitian Hamiltonian converting the optimization task into the boundary value problem. In some cases, this could be solved analytically, providing the insights into optimal control of the simplest quantum systems.

In addition, the problem of finding time-optimal solution can be presented as quantum geodesic search~\cite{Wang2015} providing a geometric interpretation to quantum brachistochrone technique. While this approach proved to be successful in several specific types of problems~\cite{Carlini2011Mar, Carlini2012Nov, Carlini2013Jan, Carlini2017Feb}, obtaining such solutions remains challenging, especially in large-scale systems with multiple degrees of freedom and many control parameters.

%Although finding solutions to a boundary-value problem for a large set of ordinary differential equations is challenging, the problem was reformulated as a quantum geodesic search~\cite{Wang2015}, providing quantum brachistochrone equations via changing a metric and demonstrating numerical applications. However, while the quantum brachistochrone method application was very fruitful in several specific types of problems~\cite{Carlini2011Mar, Carlini2012Nov, Carlini2013Jan, Carlini2017Feb}, obtaining such solutions remains a challenging problem, especially in systems with a large number of degrees of freedom (control parameters).

% \blue{ Next formulate our problem, e.g:}

\begin{figure}[t]
\centering\includegraphics[width=7cm]{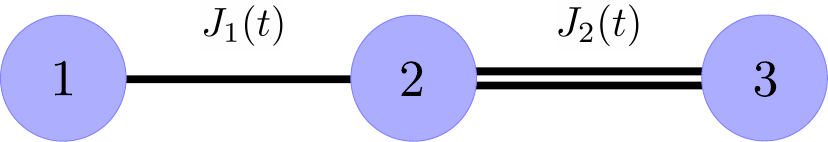}
\caption{Sketch of the three-qubit array with controllable nearest-neighbor couplings $J_1(t), J_2(t)$.}
\label{pic 1}
\end{figure}

In this Article, we illustrate quantum brachistochrone technique on a simple but instructive example. Specifically, we study time-optimal transfer of a single-particle excitation in the array of three nearest-neighbor coupled qubits (two-level systems) shown schematically in Fig.~\ref{pic 1}. We assume that initially the excitation is launched in the leftmost qubit of the array. Then, by varying the couplings $J_{1,2}$ in time such that $J_1^2+J_2^2=J_0^2=\text{const}$, we aim to achieve the fastest possible transfer of the excitation to the rightmost qubit. For simplicity, we assume that the system is non-dissipative and the eigenfrequency of the qubits is fixed.

Note that this problem is analogous to time-optimal population transfer in a three-level system where the direct transition between the first and the third levels is prohibited. Even though the solution to this problem was proposed long ago~\cite{Gottlieb1982}, its optimality was proven much later using a different technique~\cite{Boscain2002May}.

%Our goal is to provide a summary of the state-of-the-art of the brachistachrone method and demonstrate how one can apply this technique to find the time-optimal transfer of the single-particle quantum state from the left to the right edge of the array with 100\% efficiency. The transfer is performed only with a control of the couplings' amplitudes. 

% \blue{Fig.1: sketch of the 3 qubits, designation of the couplings, etc.}

% This problem does have some known solutions. One of them is stepwise switching transfer when couplings between two adjoining qubits are included in stages. Another solution is perfect transfer~\cite{perfecttransfer} with constant coupling and time-independed Hamiltonian. This case allows obtain solution with perfect efficiency of the state transfer. However both methods are not time-optimal.

% \red{
% the solution is known before 10.1103/PhysRevA.26.3713 but only later proven to be optimal: https://doi.org/10.1063/1.1465516

% and even applied for some fancy stuff
% https://journals.aps.org/prapplied/abstract/10.1103/PhysRevApplied.16.064040
% (maybe we don't need this
% }

% Mention that this problem does have a solution, couple of examples of perfect transfer. Cite earlier papers for 3-level systems.

% \commentMaxim{Need to clearly distinguish what is new and special in our main paper, which we are working on. In order not to compromise the novelty.}

\section{Summary of quantum brachistochrone method}

%Let ${\hat{A}_i}$ and ${\hat{B}_j}$ are the orthonormal bases of $\mathbb{A}$ and $\mathbb{B}$ subspaces of Hermitian space by traceless Hermitan $N \times N$ matrices with orthogonality relation  

To derive the time-optimal evolution of a quantum system, we introduce a set of $N\times N$ traceless Hermitian matrices ${\hat{A}_i}$ and ${\hat{B}_j}$ spanning the subspaces $\mathbb{A}$ and $\mathbb{B}$ and normalized by the relations $\Tr(\hat{A}_i\hat{A}_j) = \delta_{ij}, \Tr(\hat{B}_i\hat{B}_j) = \delta_{ij}, \Tr(\hat{A}_i\hat{B}_j) = 0$. We assume that the Hamiltonian can only contain the matrices from $\mathbb{A}$ subspace, while ${\hat{B}_j}$ matrices are unavailable because of the physical constraints on the system:
\begin{equation}
\hat{H} = \sum_i\alpha_i\hat{A}_i\:.
\end{equation}
Also, we assume that the norm of the Hamiltonian is bounded $||\hat{H}(t)|| = \sqrt{\Tr \hat{H}^2(t)}\leq \Delta E$. Our goal is to find such temporal variation of the Hamiltonian $\hat{H}(t)$ that the initial state $\psi (0)$ will be transferred to the final state $\psi (\tau) = \hat{U}(\tau) \psi (0)$ within the minimal possible time $\tau$, where $\hat{U}(t)$ is a unitary evolution operator satisfying Shr\"odinger equation $i \partial\hat{U}/\partial t = \hat{H}\hat{U}$. If the evolution operator is known, the Hamiltonian can be readily expressed as
\begin{equation}\label{eq:HUconnection}
\hat{H} = i\frac{\partial\hat{U}}{\partial t} \hat{U}^\dagger \:.  
\end{equation}

Obviously, the transfer time $\tau$ is inversely proportional to the bound $\Delta E$, while their product $\Delta E\,\tau$ is the dimensionless coefficient dependent on the chosen $\hat{H}(t)$ protocol. Therefore, the original problem of finding minimum possible transfer time $\tau$ for a fixed constraint $\Delta E$ is equivalent to finding the minimum possible $\Delta E$ for the prescribed transfer time $\tau=1$. This motivates the choice of the  functional~\cite{Malikis2024}
\begin{equation}
    S = \int\limits_0^\tau ||\hat{H}(t)|| \,dt + \int\limits_0^\tau \sum_k \lambda_k \mathrm{Tr}\left(\hat{B}_k\hat{H}\right) \,dt\:, 
\end{equation}
where the first term is aimed to minimize the norm of the Hamiltonian (i.e. $\Delta E$), while the second set of terms constrains the form of the Hamiltonian excluding the contribution from $\hat{B}_k$ matrices. The coefficients $\lambda_k$ are time-dependent Lagrange multipliers.

Making use of Eq.~\eqref{eq:HUconnection}, we present the target functional $S$ in the form
\begin{eqnarray}
    S = S_1 + S_2 =\int\limits_0^\tau L_T^0\,dt + i \int\limits_0^\tau \sum_k \lambda_k \mathrm{Tr}\left(\hat{B}_k\dfrac{\partial\hat{U}(t)}{\partial t}\hat{U}^\dagger(t)\right) \,dt
\end{eqnarray}
with $L_T^0 = ||\hat{H}(t)||=\sqrt{\mathrm{Tr}\left(\partial \hat{U}^\dagger/\partial t\cdot\partial \hat{U}/\partial t\right)}$. Thus the, the target functional only dependends from the evolution operator $\hat{U}(t)$ and its time derivative.

% Since the Hamiltonian is a function of the evolution operator, Eq.~\eqref{eq:HUconnection}, the desired optimal control is recovered by 

% Therefore Hamiltonian is a function of evolution operator $\hat{U}$ and it is enough to consider the variation of action only by $\hat{U}$. Note, that along the trajectory one can rescale time leading to $||\hat{H}|| = E$~\cite{Malikis2024}. To search for trajectory we consider the action in the following form:

% where $\lambda_k$ - Lagrange multipliers, a set of conditions on Hamiltonian $\Tr(\hat{B}_k\hat{H}) = 0$ is equivalent to $\hat{H} = \sum_i\alpha_i\hat{A}_i$, and $L_T^0 = ||\hat{H}(t)||=\sqrt{\mathrm{Tr}\left(\partial \hat{U}^\dagger/\partial t\cdot\partial \hat{U}/\partial t\right)}$.

Varying the functional $S$ with respect to the evolution operator, we recover
\begin{eqnarray}
    \delta S_1 &=& \dfrac{1}{2}\int\limits_0^\tau \, \dfrac{1}{L_T^0}\mathrm{Tr}\left( \dfrac{\partial \hat{U}^\dagger}{\partial t}\dfrac{\partial \delta \hat{U}}{\partial t} +  \dfrac{\partial \delta \hat{U}^\dagger}{\partial t}\dfrac{\partial \hat{U}}{\partial t} \right)dt \\ \nonumber
    &=& \dfrac{1}{L_T^0}\mathrm{Tr}\left( \dfrac{\partial \hat{U}^\dagger}{\partial t} \delta \hat{U} \right)\bigg|_0^\tau + \dfrac{1}{L_T^0}\int\limits_0^\tau \, \mathrm{Tr} \left( \left( \hat{U}^\dagger \dfrac{\partial^2 \hat{U}}{\partial t^2}\hat{U}^\dagger  + \hat{U}^\dagger\dfrac{\partial \hat{U}}{\partial t}\dfrac{\partial \hat{U}^\dagger}{\partial t}\right) \delta\hat{U}\right)dt\:,
\end{eqnarray}
where we used the identity
% $\hat{U}^\dagger\hat{U} = \mathbb{I}$, so 
$\delta \hat{U}^\dagger = - \hat{U}^\dagger\delta \hat{U}\hat{U}^\dagger$. Similarly, we compute the variation of $S_2$:

\begin{eqnarray}
    \delta S_2 &=& i\,\mathrm{Tr}\sum_k(\hat{U}^\dagger\hat{B}_k\lambda_k\delta\hat{U})\bigg|_0^\tau 
\nonumber\\&&
- i\int\limits_0^\tau \mathrm{Tr}\left( \sum\limits_k \left( \lambda_k\hat{U}^\dagger\hat{B}_k\dfrac{\partial \hat{U}}{\partial t} \hat{U}^\dagger + \hat{U}^\dagger\hat{B}_k\dfrac{\partial \lambda_k}{\partial t} +  \dfrac{\partial \hat{U}^\dagger}{\partial t}\hat{B}_k\lambda_k\right)\delta\hat{U}\right) \,dt \;,
\end{eqnarray}
where we used full derivative $\frac{\partial}{\partial t}(\lambda_k\delta\hat{U}\hat{U}^\dagger) = \frac{\partial \lambda_k}{\partial t}\delta\hat{U}\hat{U}^\dagger + \lambda_k\frac{\partial \delta\hat{U}}{\partial t}\hat{U}^\dagger + \lambda_k\delta\hat{U}\frac{\partial \hat{U}^\dagger}{\partial t}$. 

Since the initial and final states of the quantum system are fixed, $\delta \hat{U}(0) = \delta \hat{U}(1) = 0$. Moreover,  $L_T^0 = \Delta E$ along the trajectory, and thus we can rescale $L_T^0\lambda_k \rightarrow \lambda_k$. Requiring the extremum of the functional $\delta S = \delta S_1 + \delta S_2 = 0$, we obtain {\it quantum brachistochrone equation}~\cite{Carlini2007}: 
\begin{equation}\label{eq:QBE}
    \frac{d\hat{F}}{dt}+i\left[\hat{H},\hat{F}\right]=0\:,
\end{equation}
where $\hat{F}=\hat{H}+\sum_k \lambda_k\,\hat{B}_k$.
%
% \begin{align}
%     \dfrac{\partial \hat{H}}{\partial t} + \sum\limits_k \left( i\lambda_k[\hat{H},\hat{B}_k]+ \hat{B}_k\dfrac{\partial \lambda_k}{\partial t} \right) = 0
% \end{align}
Projecting this equation on the matrices $\hat{A}_m$ and $\hat{B}_n$ and taking into account their orthogonality, we recover the system
\begin{eqnarray}
    \dfrac{d\alpha_m}{dt} &=& i \sum_k \lambda_k\mathrm{Tr}\left( [\hat{A}_m, \hat{B}_k]\hat{H}\right),\label{breqs1}\\ 
    \dfrac{d\lambda_n}{dt} &=& i \sum_k \lambda_k\mathrm{Tr}\left( [\hat{B}_n, \hat{B}_k]\hat{H}\right).\label{breqs2}
\end{eqnarray}
Equations~\eqref{breqs1} define the evolution of control parameters $\alpha_m$ in the optimal scenario, while the complementary Eqs.~\eqref{breqs2} determine the change of Lagrange multipliers $\lambda_n$ in time. Notably the initial conditions for $\lambda_n$ are unknown, which makes quantum brachistochrone equations hard to solve.

Note also that in the absence of constraints on the Hamiltonian (i.e. when $\hat{B}$ matrices are absent) the time-optimal strategy is straightforward. Equation~\eqref{eq:QBE} suggests that the Hamiltonian is time-independent and should directly couple the initial and final states of the quantum system.

\section{Derivation of time-optimal evolution}

Now we apply quantum brachistochrone approach to the specific system~-- array of 3 qubits, Fig.~\ref{pic 1}. Overall, the dimensionality of the Hilbert space for such system is $2^3=8$. However, since the Hamiltonian conserves the number of excitations and since we focus on a single-particle sector, the dynamics of interest happens in the 3-dimensional subspace spanned by the three single-particle basis states. In turn, the Hamiltonian is parametrized by the two variables which are nearest-neighbor couplings $J_1$ and $J_2$.

In these conventions, $\hat{A}$ and $\hat{B}$ matrices are expressed via Gell-Mann matrices and read
\begin{align}
    &\hat{A}_1= \frac{1}{\sqrt{2}}\begin{pmatrix}
        0& 1& 0\\
        1& 0& 0\\
        0& 0& 0\\
    \end{pmatrix}, \quad
    \hat{A}_2= \frac{1}{\sqrt{2}}\begin{pmatrix}
        0& 0& 0\\
        0& 0& 1\\
        0& 1& 0\\
    \end{pmatrix}\\
    &\hat{B}_1 = \frac{1}{\sqrt{2}}\begin{pmatrix}
        0& -i& 0\\
        i& 0& 0\\
        0& 0& 0\\
    \end{pmatrix}, \quad
    \hat{B}_2 = \frac{1}{\sqrt{2}}\begin{pmatrix}
        0& 0& 1\\
        0& 0& 0\\
        1& 0& 0\\
    \end{pmatrix}, \quad
    \hat{B}_3 = \frac{1}{\sqrt{2}}\begin{pmatrix}
        0& 0& i\\
        0& 0& 0\\
        -i& 0& 0\\
    \end{pmatrix},\\
    &\hat{B}_4 = \frac{1}{\sqrt{2}}\begin{pmatrix}
        0& 0& 0\\
        0& 0& -i\\
        0& i& 0\\
    \end{pmatrix}, \quad
    \hat{B}_5 = \frac{1}{\sqrt{2}}\begin{pmatrix}
        1& 0& 0\\
        0& -1& 0\\
        0& 0& 0\\
    \end{pmatrix}, \quad
    \hat{B}_6 = \frac{1}{\sqrt{2}}\begin{pmatrix}
        1& 0& 0\\
        0& 1& 0\\
        0& 0& -2\\
    \end{pmatrix}\:,
\end{align}
while the Hamiltonian of system is presented as:
\begin{eqnarray}
    \hat{H} = \sum_{m=1}^{2}\alpha_{m} \hat{A}_{m} = \begin{pmatrix}
        0& J_1& 0\\
        J_1& 0& J_2\\
        0& J_2& 0\\
    \end{pmatrix}
\end{eqnarray}
with $J_1=\alpha_1/\sqrt{2}, J_2=\alpha_2/\sqrt{2}$. Starting from the general brachistochrone equations~\eqref{breqs1}-\eqref{breqs2}, we derive control equations for our case:
\begin{align}
\begin{cases}
&\df{J_1}=-\lambda_3\,J_2\:,\\
&\df{J_2}=\lambda_3\,J_1\:,\\
&\df{\lambda_3}=0\:.
\end{cases}
\end{align}
Interestingly, the equations for $J_1$, $J_2$ and $\lambda_3$ decouple from the rest of the system, which strongly simplifies the solution. Taking into account the constraint $J_1^2+J_2^2=J_0^2$, we derive an analytical solution:
\begin{eqnarray}\label{l3}
    \lambda_3 &=& \Omega\\ \label{j1}
    J_1 &=& J_0\cos(\Omega t +\varphi)\\ \label{j2}
    J_2 &=& J_0\sin(\Omega t + \varphi)\:,
\end{eqnarray}
where $\varphi$ is a constant phase which depends on the initial conditions. Using the obtained couplings and solving   Shr\"odinger equation we compute the components of the wave function $\ket{\psi} = (\psi_1,\psi_2,\psi_3)^T$:
\begin{eqnarray}
    \psi_1 (t) &=& -\frac{iJ_2(t)}{\Omega}A + \frac{\omega J_1 + iJ_2\Omega}{J_0^2}B_+e^{-i\omega t} - \frac{\omega J_1 - iJ_2 \Omega}{J_0^2}B_-e^{i\omega t},\\
    \psi_2 (t) &=& A + B_+e^{-i\omega t} + B_-e^{i\omega t},\\
    \psi_3 (t) &=& \frac{iJ_1(t)}{\Omega}A + \frac{\omega J_2 - iJ_1\Omega}{J_0^2}B_+e^{-i\omega t} - \frac{\omega J_2 + iJ_1 \Omega}{J_0^2}B_-e^{i\omega t},
\end{eqnarray}
where  $\omega = \sqrt{\Omega^2+J_0^2}$. Unknown integration constants $A, B_+, B_-, \Omega, \varphi$ are recovered from the boundary conditions defining eventually the transfer time $\tau$.

In our case, the initial and final states are $\ket{\psi_i} = (1,\quad 0,\quad 0)^T$, $\ket{\psi_f} = (0,\quad 0,\quad e^{i\gamma})^T$, where $\gamma$ is a global phase irrelevant for our purposes. We are interested in the fastest possible transfer of the particle from the leftmost qubit. Therefore, it is logical to maximize the coupling $J_1$ at $t=0$, while keeping $J_2(0)=0$ which yields the phase $\varphi=0$. This conclusion is also supported by numerical simulations for the different phases $\varphi$.

Combining this with the boundary conditions, we derive $\Omega=\frac{J_0}{\sqrt{3}}$ and the transfer time
\begin{equation}
    \tau = \dfrac{\sqrt{3}\pi}{2J_0} \approx 2.721/J_0\:.
\end{equation}
This provides the fastest possible transfer protocol given the nearest-neighbor couplings and the constraint on the sum of their squares.

% \begin{equation}
%    \tau = \frac{\pi}{2\Omega},\quad \Omega=\frac{J_0}{\sqrt{3}} 
% \end{equation}
% Particularly, for the initial state $\ket{\psi_i} = (1,\quad 0,\quad 0)^T$ and final state $\ket{\psi_f} = (0,\quad 0,\quad e^{i\phi})^T$, where $\phi$ - global phase, we find $\Omega = J_0/\sqrt{3}$. The $\varphi$ sets the initial value of $J_1$ and the final value of $J_2$. We need the fastest evolution of $\ket{\psi}$ to reach a time-optimal transfer. From the Shr\"odinger equation for the first wave function component $i\partial\ket{\psi_1}/\partial t = J_1\ket{\psi_2}$, one may conclude, that it is achieved by maximization of $J_1(0) = J_0$ in initial time moment. 
% This conclusion is supported by numerical calculations for different values of $\varphi\in[0,\pi/2]$. The optimal strategy is obtained for $\varphi = 0$ which allows us to calculate the transfer's minimum time.

% \begin{eqnarray}
%     \ket{\psi_i} = \begin{pmatrix}
%         1\\
%         0\\
%         0\\
%     \end{pmatrix}, \quad
%     \ket{\psi_f} = \begin{pmatrix}
%         0\\
%         0\\
%         e^{i\phi}\\
%     \end{pmatrix}
% \end{eqnarray}

\begin{figure}[h!]
\centering\includegraphics[width=\textwidth]{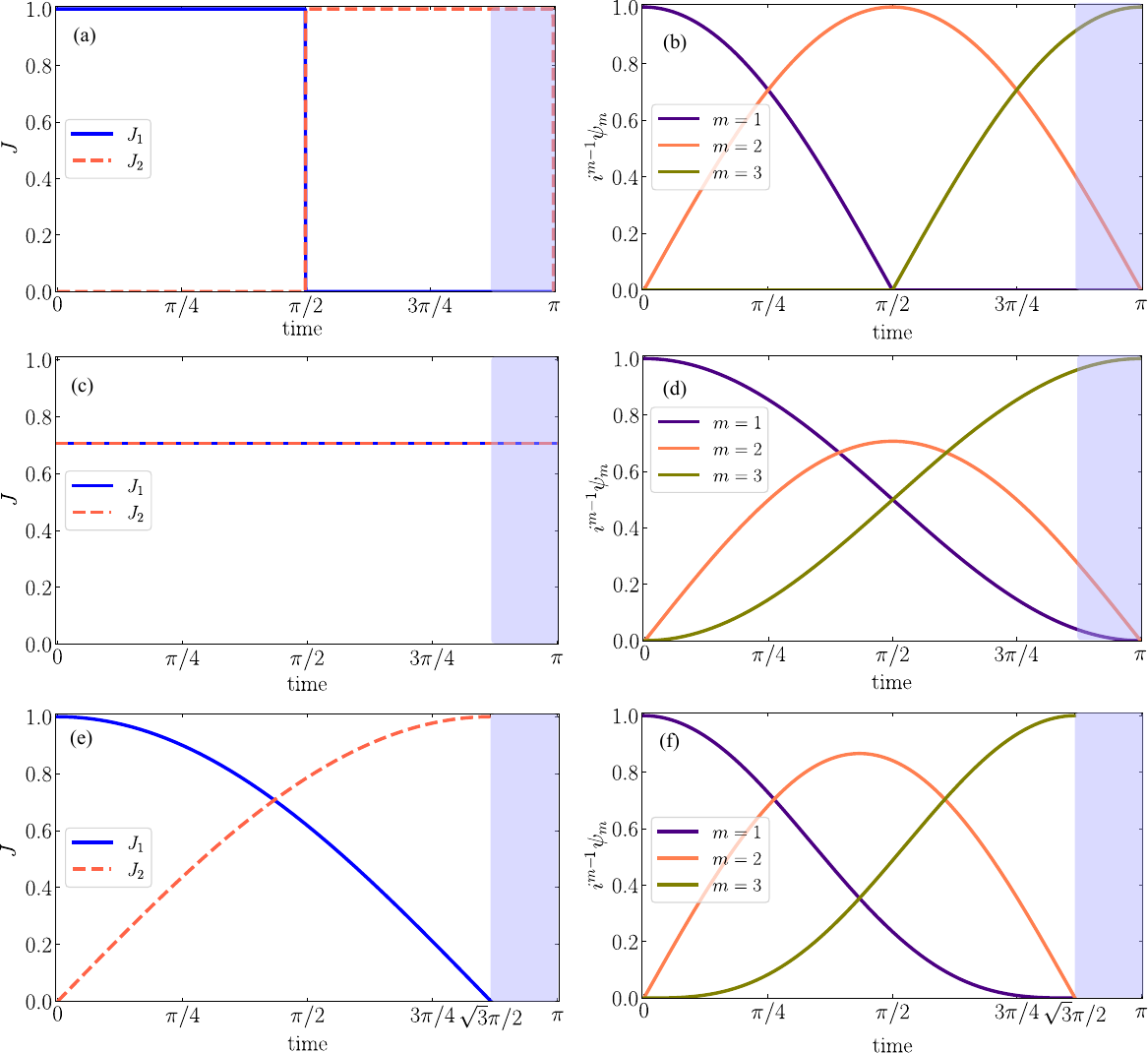}
\caption{Couplings amplitudes $J_1$ and $J_2$ for the (a) stepwise switching, (c) perfect transfer, (e) optimal transport and the calculated wave functions $i^{m-1}\psi_m$ for the (b) stepwise switching transport (d) perfect transfer (f) optimal transport. Transfer time in the scenario is 13\% less than in two other instances.}
\label{pic 2}
\end{figure}

To complete our analysis, we compare the derived protocol with the two alternative strategies also providing maximal fidelity. The first approach is a stepwise switching of the couplings [Fig.~\ref{pic 2}(a)]. In such case, $J_1(t)=J_0$ is switched on for some time until the particle moves from first qubit to the second one. Then this coupling is off and coupling $J_2$ is switched on instead. Straightforward calculation using QuSpin package~\cite{Quspin} shows that such strategy indeed provides maximal fidelity of the transfer [Fig.~\ref{pic 2}(b)]. However, the timing in this case $\tau_{st}=\pi/J_0\approx 3.142/J_0$ is  non-optimal.

Another strategy known as perfect transfer~\cite{perfecttransfer} suggests time-independent couplings both equal to $J_0/\sqrt{2}$ [Fig.~\ref{pic 2}(c)]. In this case, the particle is perfectly transferred from the first to the third site [Fig.~\ref{pic 2}(d)]. However, the timing is also non-optimal $\tau_{pt} = \pi/J_0\approx 3.142/J_0$.

These results should be compared with the calculated optimal control Eqs.~\eqref{j1}-\eqref{j2} which implies the change of the couplings according to $\cos$ and $\sin$ functions, Fig. \ref{pic 2} (e). Although the wave function's evolution shown in Fig. \ref{pic 2}(f) strongly resembles the two previous scenarios, the time of the transfer here is reduced by 13\%.

% \begin{figure}[h!]
% \centering\includegraphics[width=\textwidth]{OMEX fig 2 2.0.pdf}
% \caption{Couplings amplitudes for the (a) stepwise switching transport (c) perfect transfer (e) optimal transport and wave functions $i^{m-1}\psi_m$ for the (b) stepwise switching transport (d) perfect transfer (f) optimal transport}
% \end{figure}

% \begin{figure}[ht!]
% \centering\includegraphics[width=7cm]{opticafig1}
% \caption{Sample caption (Fig. 2, \cite{Yelin:03}).}
% \end{figure}

\section{Discussion and conclusions}
% \blue{Discuss some disadvantages and future prospects - to give more room to our main paper.}

In summary, quantum brachistochrone method is a powerful tool providing analytical insights into time-optimal control of relatively simple quantum systems. It provides an elegant solution to the two-factor optimization problems, such as finding  strategies ensuring both maximal fidelity of the transfer and the minimal transfer time.

However, optimal control of large quantum systems still poses a significant challenge, as the number of quantum brachistochrone equations grows rapidly with the system size requiring extensive computations. Given current advances in the engineering of multi-qubit quantum processors, this provides an interesting topic for further studies.

% the quantum brachistochrone technique is a powerful tool promising to provide optimal control for a wide range of applications. For example, it was used to find analytical solutions in systems with a limited number of controlled parameters, including realization optimal CNOT gate~\cite{Carlini2011Mar, Carlini2013Jan, Carlini2017Feb}, transfer of coherence~\cite{Carlini2012Nov}, and the introduction of the class of completely integrable brachistochrone problems~\cite{Malikis2024}. Based on these examples we assume that the method would be fruitful for similar non-trivial tasks with controlled time-limited processes, such as the preparation of mixed states in minimal time and second harmonic generation. \textcolor{purple}{[Kseniia: looks unclear and needs some improvement]}

% However, applying quantum brachistochrone methods for problems with a large number of controlled parameters leads to a significant complication in finding a solution, which is computationally demanding.
% Nevertheless, we anticipate that the considered problem of 100\% fidelity transfer of single-particle excitation through the 3-qubit chain in minimal time can be generalized to larger systems and more dimensional problems and is left for further research.

\begin{backmatter}
\bmsection{Funding} Theoretical models were supported by Priority 2030 Federal Academic Leadership Program. Numerical simulations were supported by the Russian Science Foundation, grant No.~24-72-10069. M.A.G. acknowledges partial support by the Foundation for the Advancement of Theoretical Physics and Mathematics ``Basis''.

% \bmsection{Acknowledgments}
% The section title should not follow the numbering scheme of the body of the paper. Additional information crediting individuals who contributed to the work being reported, clarifying who received funding from a particular source, or other information that does not fit the criteria for the funding block may also be included; for example, ``K. Flockhart thanks the National Science Foundation for help identifying collaborators for this work.'' 

\bmsection{Disclosures}
The authors declare no conflicts of interest

\end{backmatter}

\bibliography{sample}

%%%%%%%%%% If preparing manually:
% \begin{thebibliography}{1}
% \newcommand{\enquote}[1]{``#1''}

% \bibitem{Zhang:14}
% Y.~Zhang, S.~Qiao, L.~Sun, Q.~W. Shi, W.~Huang, L.~Li, and Z.~Yang,
%   \enquote{Photoinduced active terahertz metamaterials with nanostructured
%   vanadium dioxide film deposited by sol-gel method,}
%   {\protect\JournalTitle{Optics Express}} \textbf{22}, 11070--11078 (2014).

% \bibitem{Optica}
% {Optica}, \enquote{{Optica Publishing Group},}
%   \url{http://www.opg.optica.org}.

% \bibitem{FORSTER2007}
% P.~Forster, V.~Ramaswamy, P.~Artaxo, T.~Bernsten, R.~Betts, D.~Fahey,
%   J.~Haywood, J.~Lean, D.~Lowe, G.~Myhre, J.~Nganga, R.~Prinn, G.~Raga,
%   M.~Schulz, and R.~V. Dorland, \enquote{Changes in atmospheric consituents and
%   in radiative forcing,} in \enquote{Climate Change 2007: The Physical Science
%   Basis. Contribution of Working Group 1 to the Fourth assesment report of
%   Intergovernmental Panel on Climate Change,}  S.~Solomon, D.~Qin, M.~Manning,
%   Z.~Chen, M.~Marquis, K.~B. Averyt, M.~Tignor, and H.~L. Miler, eds.
%   (Cambridge University Press, 2007).

% \end{thebibliography}

\end{document}